# Unconventional superconductivity

# in CeCoIn$_5$ with magnetic texture

# and orbital quantization

H.A. Radovan*, N.A. Fortune†, T.P. Murphy*, S.T. Hannahs*, E.C. Palm*, S.W. Tozer* & D. Hall*[1]

* *NHMFL, Florida State University, Tallahassee, Florida 32310, USA*

† *Department of Physics, Smith College, Northampton, Massachusetts 01063, USA*

[1]Present address: American Physical Society, One Research Road, Box 9000, Ridge, New York 11961

**A sufficiently high magnetic field applied to a superconductor will act on both the charge and the spin of the individual electrons breaking up the Cooper pairs. If the spin effect dominates, the superconducting state can develop a texture before eventually entering the normal state with increasing magnetic field. This spatially varying superconducting state is a periodic array of magnetic walls separated by superconducting regions. Known as the Fulde-Ferrell-Larkin-Ovchinnikov (FFLO) phase, it was predicted in 1964.[1,2] We report heat capacity measurements on the heavy fermion superconductor, CeCoIn$_5$, which reveal a second phase transition within the superconducting state, clear evidence of the FFLO phase. We also report magnetization measurements that display a cascade of first order phase transitions within the FFLO region. Each transition indicates an increase in orbital momentum**

**of the superconducting order parameter and corresponds to a specific Landau level vortex state comprised of multiquanta vortices.[3,4] The experimental realization of the FFLO state provides a new opportunity to study the symbiosis of magnetism and superconductivity, two states of matter once thought to be mutually exclusive.**

In the standard BCS model, orbital effects limit superconductivity, which is destroyed when the kinetic energy of the charges exceeds the condensation energy of the Cooper pairs. The expected zero temperature critical field is given by Werthamer-Helfand-Hohenberg formula $B_{orb}(T=0) = 0.7T_c|dB_{c2}/dT|_{Tc}$, where $T_c$ is the critical temperature at zero field.[5] In the paramagnetic or Pauli limit, superconductivity is destroyed when the polarization energy of the spins exceeds the condensation energy due to partial alignment of the spins in field. Here, the zero temperature critical field is given by the Clogston limit, $B_p(T=0) = \Delta_0/\sqrt{2}\mu_B = 1.85T_c$ (Kelvin), where $\Delta_0$ is the energy gap, $\mu_B$ is the Bohr magneton, and $T_c$ is the critical temperature.[6] The ratio of these limits is known as the Maki parameter, $\alpha \equiv \sqrt{2}B_{orb}/B_p$, and indicates whether a material is orbitally or paramagnetically limited.[7]

Fulde & Ferrell and, independently, Larkin & Ovchinnikov predicted a transition within the superconducting state between the paramagnetic and orbital limits.[1,2] The superconducting order parameter (Cooper pair condensate) lowers its free energy by becoming spatially inhomogenous. This FFLO state consists of a spatially modulated order parameter and an array of Bloch-type walls of broken Cooper pairs with polarized spins. The FFLO phase is predicted to be stable at temperatures below $\approx T_c/2$ with a field dependent Bloch wall separation on the order of the coherence length, $\xi$. $T_{FFLO}$ cannot exceed $0.56T_c$, but the influence of material parameters may result in lower values.[8,9] Within the BCS picture, this state corresponds to Cooper pairs having a finite center-of-mass momentum due to coupling across the Zeeman-energy-split Fermi surface for up- and down-spin electrons. The FFLO phase is an example of spontaneous spatial symmetry breaking.

The FFLO scenario was generalized in 1996 by taking into account both orbital and paramagnetic effects.[3,4,10,11] An experimental realization, in this case, would be a quasi-2D superconductor in an applied magnetic field slightly tilted away from the conducting planes. The field can be resolved into components normal and parallel to the conducting planes with the components being subject to orbital or paramagnetic limits, respectively. With increasing magnetic field and at a fixed temperature, $T < T_{FFLO}$, a cascade of first order transitions is expected to occur within the FFLO wedge before eventually reaching $H_{c2}$. The angle between the external field and the superconducting planes determines the highest Landau level orbital quantum number, $m$, and hence the number of crossed phase transition lines, $H_m$. The physical FFLO-to-normal state boundary, $H_{c2}(T)$, is the envelope of all $H_m(T)$ lines. Schematically shown in figure 1 are the upper critical field, $H_{c2}(T)$, the emerging spatially varying FFLO phase at low temperatures and high magnetic fields, and a series of dashed lines within the FFLO state denoting sub-phases with higher orbital momentum. Theory predicts a jump in the magnetization or critical current when entering each sub-phase $m$.[3,4] This scenario should be realized[3] for a Maki parameter, $\alpha > 9$. To summarize, in the pure orbital limit, the energetically favorable solution is the lowest

Landau level (m ≡ 0). In the pure paramagnetic case, the emerging state is the FFLO phase, while a mixture of both yields new exotic states with center-of-mass *and* orbital momenta.

Stringent requirements are imposed on the material for it to be within the paramagnetic limit: a low $T_c$; a high effective electron mass, m*, indicated by a high critical field slope at $T_c$; and a mean free path, ℓ, of the electrons that far exceeds the coherence length, ℓ $\gg \xi$, putting the material in the clean limit. Two candidate classes of materials are the heavy fermion (HF) and low-dimensional molecular superconductors. Our measurements were performed on single crystals of the HF superconductor, $CeCoIn_5$, with $T_c = 2.3$ K,[13] an initial in-plane field slope of $|dH_{c2}/dT|_{Tc} = 24$ T/K[14], and ℓ$/\xi = 14$.[15] With a Maki parameter of $\alpha \approx 13$, $CeCoIn_5$ is clearly in the paramagnetic limit, making possible the observation of the FFLO state.

$CeCoIn_5$ crystallizes in a tetragonal structure with alternating layers of $CeIn_3$ and $CoIn_2$. It belongs to the isostructural group of $CeMIn_5$ (M = Co, Ir, Rh). These materials have attracted considerable attention due to the interplay of superconductivity and magnetism. $CeRhIn_5$ has an anti-ferromagnetic (AF) ground state with pressure induced superconductivity above 16 kbar,[16] while $CeIrIn_5$ and $CeCoIn_5$ are ambient pressure superconductors. Based on these findings, the existence of a quantum critical point (QCP) has been proposed where the ground state becomes superconducting when anti-ferromagnetism is suppressed to T = 0 K by external or chemical pressure.[17-20] The proximity to an AF instability also opens up the possibility for spin-fluctuation mediated superconductivity, which has been shown to be more robust in 2D than in 3D.[21] For our experiments, we used crystal platelets with approximate dimensions of 1 mm × 1 mm × 0.1 mm and masses on the order of 0.25 mg. The single crystals were synthesized from an In flux by combining stoichiometric amounts of Ce and Co. Excess In was removed during preparation with a centrifuge and afterwards by etching in a dilute HCl/HNO$_3$ aqueous solution[13].

The main panel of figure 2 shows measurements of heat capacity as function of applied field, C vs. H, for two field orientations with respect to the conducting planes at T = 250 mK (T/T$_c$ = 0.12). At small angles, when the magnetic field is near the plane parallel orientation (θ = 0°), two phase transitions are present: a lower-field second-order phase transition within the superconducting state plus a higher field first-order phase transition at H$_{c2}$. We find that this lower-field transition exists only within ±10° of the plane parallel field orientation. At larger angles (θ = 30°), there is a single transition at H$_{c2}$. This transition is still first order, as can be seen from the magnetocaloric data for θ = 30°. This unexpected result (one would assume a second order transition in the orbital limit) is seen by various groups, but not fully understood at this time.[24,25]

The heat capacity measurements are readily explained within the FFLO scenario. The quasi-2D nature of the crystal inhibits orbital motion of the Cooper pairs in the parallel field orientation. With increasing angle, the orbital component begins to dominate and prevents the non-uniform state from appearing. The higher order, lower field transition then marks the uniform superconducting-to-FFLO state boundary. Calculations indeed predict this transition to be of second order.[26]  Our results provide the first calorimetric evidence for the FFLO state.

Figure 3 shows torque magnetometry measurements as a function of an applied magnetic field.  The main panel comprises up sweeps for five different field orientations between 20° out of plane and plane parallel.  Shown are magnetization data versus normalized applied field, $H/H_{c2}$.  The 20° curve displays a smooth increase in magnetization, much as the 30° curve in the inset.  A series of steps appears at 10° out of plane, starting above $0.85H_{c2}$ before the superconducting-to-normal transition takes place.

As the angle is decreased the steps become more pronounced and their number is reduced,

until in the pure paramagnetic limit at 0° orientation these features disappear.  It is

important to note that identical behavior is seen for the down sweep in field (not shown for

clarity), as well as for all field angles on either side of the field parallel orientation.  This

cascade signature is seen only within ±10° of parallel, the same angular orientation where

the lower FFLO phase transition line was detected.  This cascade is a direct consequence of

the spatially varying superconducting order parameter promoting higher orbital solutions.[3,4]

As discussed above, higher Landau level states of the order parameter correspond to multiquanta vortex states.  This suggests that within the FFLO state, taking into account finite orbital effects, the usual Abrikosov vortices are replaced by multiquanta vortices carrying flux $m\Phi_0$.

Figure 4 presents the experimentally determined field-temperature phase diagram for the field orientation parallel to the superconducting planes as derived from magnetization and heat capacity data for CeCoIn$_5$. We have included resistance measurements of Ref. 29 to show consistency with our data and to extend H$_{c2}$ to zero field. For temperatures below approximately 0.35 K, two distinct phase transitions emerge. We identify the temperature T ≈ 0.35 K as the onset of the FFLO state in CeCoIn$_5$. The second order, lower field transition (solid circles) marks the crossover from the uniform superconducting state to the spatially varying FFLO state, while the first order, upper transition (open symbols) identifies the FFLO-to-normal state transition H$_{c2}$(T). A first order phase transition for temperatures above 0.35 K separates the uniform superconducting state from the normal state. It becomes second order (solid symbols) for temperatures above approximately 1.3 K. Interestingly, the change in the order of the upper critical field transition coincides with the appearance of a metamagnetic transition (bowties) as seen in Ref. 29 for field aligned parallel to the planes. It is important to note that the most recent magnetotransport measurements on CeCoIn$_5$ reveal a similar line within the normal state region of the H-T phase diagram for fields applied perpendicular to the planes.[30] Thus, the metamagnetic transition and the change in criticality of H$_{c2}$ shown in Fig. 4 are, unlike the appearance of the FFLO state, independent of field orientation.

Our measurements of a series of first order transitions in the magnetization, correlated with the appearance in heat capacity measurements of an additional phase transition within the superconducting state, provides the first direct thermodynamic evidence for the existence of the FFLO state. These observations confirm a four decade old theory and open up a new possibility to study the interaction of magnetism and superconductivity, two states of matter once thought to be mutually exclusive. In this case, however, the formation of magnetic texture in form of the spin aligned Bloch walls effectively lowers the free energy of the system, making it more stable. The realization of the FFLO state in this reduced dimensional system may shed light on magnetically mediated electron pairing and the superconducting mechanism of other layered systems such as high-temperature superconductors and molecular superconductors.

[1]Fulde, P. & Ferrell, R.A. Superconductivity in a strong spin-exchange field. *Phys. Rev.* **135**, A550-A563 (1964).

[2]Larkin, A.I. & Ovchinnikov, Y.N. Inhomogeneous state of superconductors. *Sov. Phys. JETP* **20**, 762-769 (1965).

[3]Buzdin, A.I. & Brison, J.P. New solutions for the superconducting order parameter in a high magnetic field. *Phys. Lett. A* **218**, 359-366 (1996).

[4]Buzdin, A.I. & Brison, J.P. Non-uniform state in 2D superconductors. *Europhys. Lett.* **35**, 707-712 (1996).

[5]Werthamer, N.R., Helfand, E. & Hohenberg, P.C. Temperature and purity dependence of the superconducting critical field, $H_{c2}$. III. Electron spin and spin-orbit effects. *Phys. Rev.* **147**, 295-302 (1966).

[6]Clogston, A.M. Upper limit for critical field in hard superconductors. *Phys. Rev. Lett.* **9**, 266-267 (1962).

[7]Maki, K. Effect of Pauli paramagnetism on magnetic properties of high-field superconductors, *Phys. Rev.* **148**, 362-369 (1966).

[8]Norman, M.R. Existence of the FFLO state in superconducting $UPd_2Al_3$. *Phys. Rev. Lett. (Comment)* **71**, 3391 (1993).

[9]Agterberg, D.F. & Yang K. The effect of impurities on Fulde-Ferrell-Larkin-Ovchinnikov superconductors. *J. Phys.: Condens. Matter* **13**, 9259-9270 (2001).

[10]Shimahara, H. & Rainer, D. Crossover from vortex states to the Fulde-Ferrell-Larkin-Ovchinnikov state in two-dimensional s- and d-wave superconductors. *J. Phys. Soc. Jap.* **66**, 3591-3599 (1997).

[11]Houzet, M. & Buzdin, A. Structure of the vortex lattice in the Fulde-Ferrell-Larkin-Ovchinnikov state. *Phys. Rev. B* **63**, 184521-1-184521-5 (2001).

[12]Burkhardt, H. & Rainer, D. Fulde-Ferrell-Larkin-Ovchinnikov state in layered superconductors. *Ann. Physik* **3**, 181-194 (1994).

[13]Petrovic, C. *et al.* Heavy-fermion superconductivity in CeCoIn$_5$ at 2.3 K. *J. Phys.: Condens. Matter* **13**, L337-L342 (2001).

[14]Ikeda S. *et al.* Unconventional superconductivity in CeCoIn$_5$ studied by the specific heat and magnetization measurements. *J. Phys. Soc. Jap.* **70**, 2248-2251 (2001).

[15]Movshovich, R *et al.* Unconventional superconductivity in CeIrIn$_5$ and CeCoIn$_5$: Specific heat and thermal conductivity studies. *Phys. Rev. Lett.* **86**, 5152-5155 (2001).

[16]Hegger, H *et al.* Pressure-induced superconductivity in quasi-2D CeRhIn$_5$. *Phys. Rev. Lett.* **84**, 4986-4989 (2000).

[17]Sachdev, S. Quantum criticality: Competing ground states in low dimensions. *Science* **288**, 475-480 (2000).

[18]Nicklas, M. *et al.* Response of the heavy-fermion superconductor CeCoIn$_5$ to pressure: Roles of dimensionality and proximity to a quantum-critical point. *J. Phys.: Condens. Matter* **13**, L905-L912 (2001).

[19]Sidorov, V.A. *et al.* Superconductivity and quantum criticality in CeCoIn$_5$. *Phys. Rev. Lett.* **89**, 157004-1-157004-4 (2002).

[20]Kawasaki, S. *et al.* Pressure-temperature phase diagram of antiferromagnetism and superconductivity in CeRhIn$_5$ and CeIn$_3$: $^{115}$In-NQR study under pressure. *Phys. Rev. B* **65**, 020504-1 to 020504-4 (2001).

[21]Monthoux, P. & Lonzarich, G.G. Magnetically mediated superconductivity in quasi-two and three dimensions. *Phys. Rev. B* **63**, 054529-1-054529-10 (2001).

[22]Hannahs, S.T. & Fortune, N.A. Heat capacity cell for angular measurements in high magnetic fields. *Physica B*, April (2003), Proceedings of the 23$^{rd}$ International Conference on Low-Temperature Physics (LT23), Hiroshima, Japan, 20-27 Aug 2002.

[23]Fortune, N.A. *et al.* High magnetic field corrections to resistance thermometers for low temperature calorimetry. *Rev. Sci. Instrum.* **71,** 3825-3830 (2000).

[24]Tayama, T. *et al.* Unconventional heavy-fermion superconductor CeCoIn5: dc magnetization study at temperatures down to 50 mK. *Phys. Rev. B* **65**, 180504-1 to 180504-4 (2002).

[25]Bianchi, A. *et al.* First-order superconducting phase transition in CeCoIn$_5$. *Phys. Rev. Lett.* **89**, 137002-1 to 137002-4 (2002).

[26]Matsuo, S., Higashitani, S., Nagato, Y. & Nagai, K. Phase diagram of the Fulde-Ferrell-Larkin-Ovchinnikov state in a three-dimensional superconductor. *J. Phys. Soc. Jap.* **67**, 280-289 (1998).

[27]Zieve, R.J. *et al.* Vortex avalanches at one thousandth the superconducting transition temperature. *Phys. Rev. B* **53**, 11849-11854 (1996).

[28]Mola, M.M., Hill, S., Qualls, J.S. & Brooks, J.S. Magneto-thermal instabilities in an organic superconductor. *Int. J. Mod. Phys. B* **15**, 3353-3356 (2001).

[29]T.P. Murphy *at al.* Anomalous superconductivity and field-induced magnetism in CeCoIn$_5$. *Phys. Rev. B* **65**, 100514-1 to 100514-4 (2002).

[30]Paglione, J. *at al.* Field-induced fermi liquid state in CeCoIn$_5$. *Cond-mat*/0212502, submitted to *Phys. Rev. Lett.*

**Acknowledgements.** This work was supported by the National Science Foundation, the State of Florida, and NHMFL Visiting Scientist and In-House Research Programs. We would like to thank J. Sarrao (MST-10, LANL) for providing the samples. The technical assistance of V. Williams, D. McIntosh, J. Farrell, and J. Kosakowski was greatly appreciated.

Correspondence should be addressed to H.A.R. (e-mail: radovan@magnet.fsu.edu).

Figure 1. Qualitative phase diagram with uniform superconducting (U), normal (N), and FFLO phases. The inset shows the spatial modulation of the order parameter, $|\Psi|^2$, along the applied field, H, within the non-uniform FFLO phase. Taken

together, the nodes of adjacent vortices correspond to the spin-aligned Bloch walls extending perpendicular to the applied field.  The exact temperature of emergence, $T_{FFLO}$, and the temperature dependence of the uniform superconducting-to-FFLO state transition line depend on the microscopic parameters of the material, such as Fermi surface geometry and spin exchange interactions.[12]  Dashed lines within the FFLO phase illustrate higher Landau level states of the spatially varying order parameter with orbital momentum, $m$.  The solution of the linearized gap equation corresponds to the Schrödinger equation with a field term.[3,4]  The resulting eigenfunctions for the order parameter are $\Delta(r) \sim \exp(-im\varphi)\exp(-\rho^2/2)\rho^m\exp(iQz)$, with orbital quantum number, $m$, polar angle, $\varphi$, dimensionless coordinate, $\rho = r(2\pi eH/hc)^{1/2}$, and magnitude of the FFLO wave vector Q directed along the applied field H in the z direction.  This general solution is a set of Landau level functions possessing quantized orbital momentum, m, and $2\pi m$ phase change around r = 0.  The spacing of these transitions is not simply periodic in 1/B as expected for general quantum oscillations, but rather depends on the energies of different Landau level vortex states, a problem that has not been solved.[3]  The flux line lattice is the superposition of all solutions, corresponding to multiquanta vortex states instead of the usual Abrikosov vortex lattice.

Figure 2.  Heat capacity and magnetocaloric measurements of $CeCoIn_5$ vs. field at T = 250 mK.  This data was collected using a custom rotatable calorimeter placed in the mixing chamber of a top loading dilution refrigerator.[22]  The temperature was held constant by actively compensating for the magnetoresistance of the temperature sensors while continuously sweeping field at 0.003 T/min.[23]  Heat

capacity for increasing and decreasing field is shown at θ = 0° and for decreasing field at θ = 3°. The 3° down sweep has been vertically offset and the 3° up sweep has been omitted for clarity. Arrows indicate the direction of the field sweeps. At small angles, when the magnetic field is near the plane parallel orientation (θ = 0°), two phase transitions are present: a lower-field second-order phase transition between the uniform and FFLO superconducting states plus a higher field first-order phase transition at $H_{c2}$ between the FFLO and normal states. The lower-field FFLO transition is only present for θ ≤ 10°. The inset shows magnetocaloric measurements at the same temperature and a higher sweep rate of 0.03 T/min for θ = 0° and θ = 30°. These field sweeps show the absorption and release of latent heat at $H_{c2}$ for increasing and decreasing fields, respectively, confirming the first-order nature of the transition at $H_{c2}$. The absence of any latent heat for the θ = 0° data at the position marked by the arrow confirms that the FFLO transition is second order. The $H_{c2}$ phase transition for larger angles is still first order, as can be seen from the θ = 30° data.

Figure 3. Landau level quantization within the FFLO state. Magnetization vs. field at various field angles with respect to the conducting planes at T = 25 mK. The inset shows the cantilever signal for a complete field sweep cycle at 30° with a first order transition visible at 7 T. The measured torque signal, $\tau \sim M \times H$, is a voltage proportional to the gap between the flexible sample plate and a fixed reference plate. The total gap separation is measured as a capacitance using a precision capacitance bridge. The curves in the main panel are collapsed at $0.85H_{c2}$ for clarity. At large angles the magnetization increases smoothly toward $H_{c2}$, while within ±10° of the parallel orientation distinct jumps appear. These jumps represent transitions of the order parameter into higher Landau level states and are a direct consequence of the modulated FFLO phase.[3,4] We have observed these steps on two different samples using a metal BeCu (13 µm thick) and a metal MP35N (8 µm thick) cantilever which rules out sample and cantilever artifacts. We have also excluded flux avalanches as the source of these jumps. Upon repeated field sweep cycles, the jumps appear at exactly the same values of the applied field and are of uniform step size, whereas flux avalanches show a distribution of jump sizes.[27,28] No steps in the magnetization were observed at temperatures exceeding $T_{FFLO}$, confirming their non-systematic origin.

Figure 4. Experimental H-T phase diagram for $CeCoIn_5$. Transitions were obtained from heat capacity (circles), magnetization (squares & bowties) and transport (diamonds) measurements. Open symbols denote first order phase transitions; solid symbols represent second order phase transitions. The external field is applied parallel to the conducting planes. Two distinct phase transitions emerge below $T_{FFLO} \approx 0.35$ K. The lower, second order phase transition (solid circles) corresponds to the crossover from the uniform superconducting state to the spatially modulated FFLO state, while the upper boundary (open circles) corresponds to a first order phase transition into the normal state. The first order character of the transition extends beyond the tricritical point at $T_{FFLO}$ up to approximately 1.3 K. The bowties indicate a metamagnetic transition that merges with $H_{c2}$ at 1.4 K.[29] This specific temperature marks a change in $H_{c2}(T)$ from a first order to a second order phase transition with increasing temperature. The inset

shows data below 0.6 K on an expanded scale with the uniform (U) BCS superconducting state and the normal (N) state above $H_{c2}(T)$. The decrease in the lower transition field with decreasing temperature indicates a dominance of negative (anti-ferromagnetic) exchange interactions according to extensive calculations[12]. Proximity to an anti-ferromagnetic quantum critical point should favor this type of interaction. Within the FFLO phase, higher Landau level solutions for the order parameter can occur when both orbital and spin effects play a role. This is seen in the magnetization measurements as a cascade of first order phase transitions at sub-phases with higher orbital momentum (see figure 1).

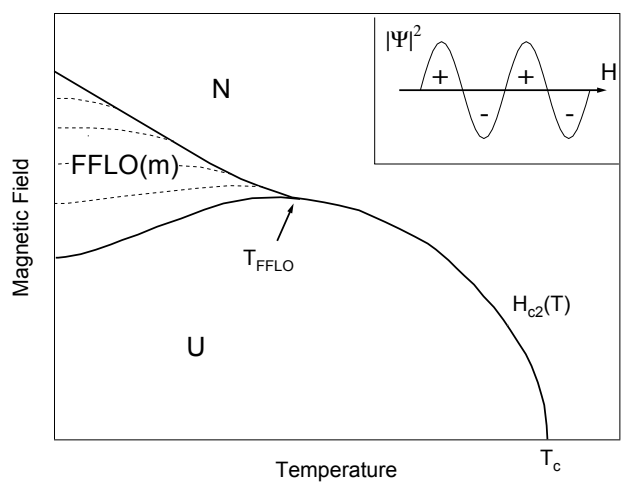

Fig. 1 (Radovan)

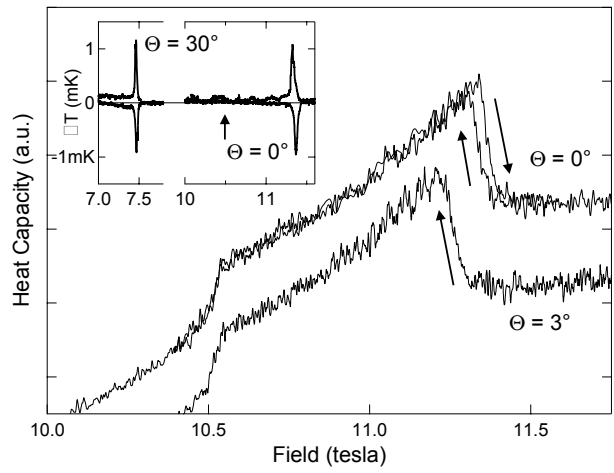

Fig. 2 (Radovan)

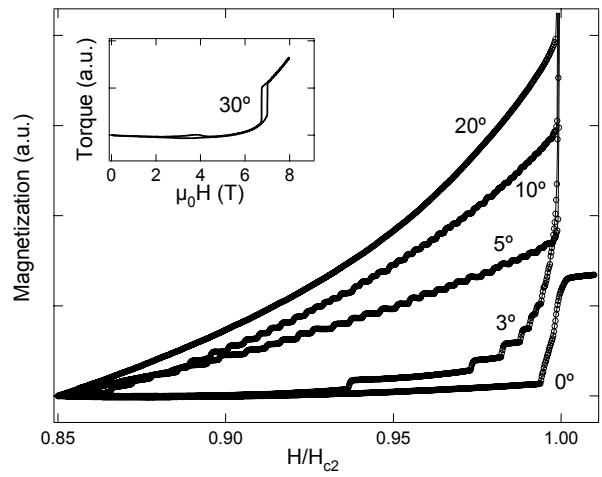

Fig. 3 (Radovan)

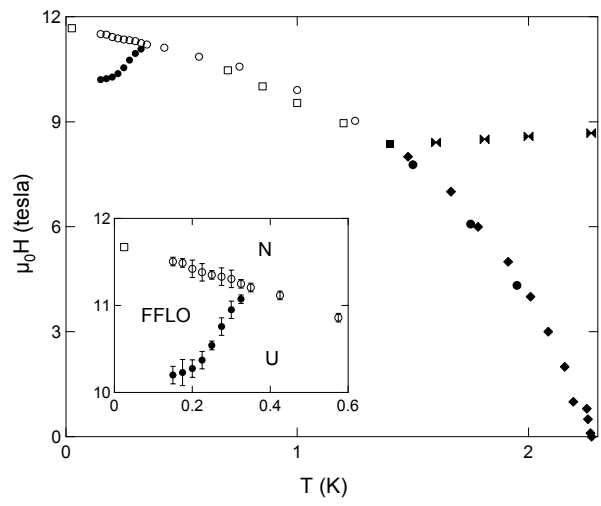

Fig. 4 (Radovan)